\documentclass[12pt,fleqn]{article}
\usepackage{amsfonts,amssymb,latexsym,cite}
\usepackage{epsf,graphicx,color}
\usepackage{epstopdf}

\def\onecol{\onecolumn \mathindent 2em}

\def\noi{\noindent}

\newcommand{\Title}[1]{\noi {{\Large\bf #1}}\\[1ex]}

\def\Aunames#1{\noi{\large\bf #1}}
\def\auth#1{${}^{#1}$}
\def\Addresses#1{\medskip\noi \protect
    \begin{description}\itemsep -3pt {\it #1} \end{description}}
\def\addr#1#2{\item[${}^{#1}$]{\it #2}}

\newcommand{\Abstract}[1]{\vskip 2mm \begin{center}
        \parbox{16.4cm}{\small\noi #1} \end{center}\medskip}

\def\email#1#2{\footnotetext[#1]{e-mail: #2}\addtocounter{footnote}{1}}

\def\nq{\hspace*{-1em}}
\def\nqq{\hspace*{-2em}}
\def\nhq{\hspace*{-0.5em}}

\def\cm{\hspace*{1cm}}
\def\inch{\hspace*{1in}}

\def\ten#1{\mbox{$\times 10^{#1}$}}
\def\deg{\mbox{${}^\circ$}}                     

\def\Jl#1#2{#1 {\bf #2},\ }

\def\ApJ#1 {\Jl{Astroph. J.}{#1}}
\def\CQG#1 {\Jl{Class. Quantum Grav.}{#1}}
\def\DAN#1 {\Jl{Dokl. AN SSSR}{#1}}
\def\GC#1 {\Jl{Grav. \& Cosmol.}{#1}}
\def\GRG#1 {\Jl{Gen. Rel. Grav.}{#1}}
\def\JETF#1 {\Jl{Zh. Eksp. Teor. Fiz.}{#1}}
\def\JETP#1 {\Jl{Sov. Phys. JETP}{#1}}
\def\JHEP#1 {\Jl{JHEP}{#1}}
\def\JMP#1 {\Jl{J. Math. Phys.}{#1}}
\def\NPB#1 {\Jl{Nucl. Phys.}{B\ #1}}
\def\NP#1 {\Jl{Nucl. Phys.}{#1}}
\def\PLA#1 {\Jl{Phys. Lett.}{#1A}}
\def\PLB#1 {\Jl{Phys. Lett.}{#1B}}
\def\PRD#1 {\Jl{Phys. Rev.}{D\ #1}}
\def\PRL#1 {\Jl{Phys. Rev. Lett.}{#1}}

\def\al{&\nhq}
\def\lal{&&\nqq {}}
\def\eq{Eq.\,}
\def\eqs{Eqs.\,}
\def\beq{\begin{equation}}
\def\eeq{\end{equation}}
\def\bear{\begin{eqnarray}}
\def\bearr{\begin{eqnarray} \lal}
\def\ear{\end{eqnarray}}
\def\earn{\nonumber \end{eqnarray}}
\def\nn{\nonumber\\ {}}
\def\nnv{\nonumber\\[5pt] {}}
\def\nnn{\nonumber\\ \lal }

\def\yy{\\[5pt] {}}
\def\yyy{\\[5pt] \lal }
\def\eql{\al =\al}

\def\dst{\displaystyle}
\def\tst{\textstyle}
\def\fracd#1#2{{\dst\frac{#1}{#2}}}
\def\fract#1#2{{\tst\frac{#1}{#2}}}
\def\Half{{\fracd{1}{2}}}
\def\half{{\fract{1}{2}}}

\def\e{{\,\rm e}}
\def\d{\partial}

\def\sign{\mathop{\rm sign}\nolimits}

\def\const{{\rm const}}
\def\eps{\varepsilon}
\def\ep{\epsilon}

\def\then{\ \Rightarrow\ }

\def\mn{_{\mu\nu}}
\def\MN{^{\mu\nu}}
\def\mN{_\mu^\nu}
\def\nM{_\nu^\mu}
\def\cK{{\cal K}}
\def\cV{{\cal V}}

\def\kappa{\varkappa}

\def\wt{\widetilde}
\def\tg{{\wt g}}
\def\tR{{\wt R}}

\def\oR{{\overline R}}

\def\sss{\scriptscriptstyle}

\def\mD{m_{\sss D}}

\begin{document}
\thispagestyle{empty}
\onecol

\Title{Nonlinear multidimensional gravity\yy
    and the Australian dipole}

\Aunames{\small
     K.A. Bronnikov\auth{a,b,1},
     V.N. Melnikov\auth{a,b,2},
     S.G. Rubin\auth{c,3},
     and I.V. Svadkovsky\auth{c,4}}

\Addresses{ \small
\addr a {Center for Gravitation and Fundamental Metrology, VNIIMS,
            46 Ozyornaya St., Moscow 119361, Russia}
\addr b {Institute of Gravitation and Cosmology, PFUR,
            6 Miklukho-Maklaya St., Moscow 117198, Russia}
\addr c {National Research Nuclear University ``MEPhI'',
     Kashirskoe sh. 31, Moscow, 115409 Russia}
      }

\Abstract
  {The existing observational data on possible variations of fundamental
   physical constants (FPC) confirm more or less confidently only a
   variability of the fine structure constant $\alpha$ in space and time.
   A model construction method is described, where variations of $\alpha$
   and other FPCs (including the gravitational constant $G$) follow from
   the dynamics of extra space-time dimensions in the framework of
   curvature-nonlinear multidimensional theories of gravity. An advantage
   of this method is a unified approach to variations of different FPCs.
   A particular model explaining the observable variations of $\alpha$
   in space and time has been constructed. It comprises a FRW cosmology
   with accelerated expansion, perturbed due to slightly inhomogeneous
   initial data.
   }

\email 1 {kb20@yandex.ru}
\email 2 {melnikov@phys.msu.ru}
\email 3 {sergeirubin@list.ru}
\email 4 {igor\_svadkovsky@mail.ru}

\section{Introduction}

  The problem of possible variations of the fundamental physical constants
  (FPC) in time and space is one of the most challenging problems of modern
  physics, directly related to the central problem of unification of all
  interations. It traces back to Dirac's and Eddington's famous papers of
  the 1930s and since then gains much attention in both theoretical and
  experimental studies.

  However, to date, a variability of only one FPC has been revealed by
  observations more or less confidently, it is the fine structure constant
  $\alpha$. The analysis of absorption spectra of various ions in the
  radiation of distant quasars, performed in the recent years (above all,
  from the data obtained at the Keck telescope on the Hawayian islands),
  has led to a conclusion that $\alpha$ is changing with time, so that in
  the past it was slightly smaller than now (the relative change
  $\delta\alpha/\alpha$ is about $10^{-5}$ \cite{webb1}). In 2010, an
  analysis of new data obtained at the VLT (Very Large Telescope), located
  in Chile, and their comparison with the Keck data led to a conclusion on
  spatial variations of $\alpha$, i.e., on its dependence on the direction
  of observations. The VLT observations in the Southern part of the
  celestial sphere gave values of the parameter $\alpha$ in the past
  slightly larger than now. This anisotropy has a dipole nature
  \cite{webb2,webb3} and has been termed ``the Australian dipole''
  \cite{beren1}. The dipole axis is located at a declination of
  $-61 \pm 9\deg$ and at a right ascention of $17.3 \pm 0,6$ hours. The
  deflection of $\alpha$ value at an arbitrary point $r$ of space from its
  modern value $\alpha_0$, measured on Earth, is
\beq                                \label{dipole}
    \delta\alpha/\alpha_0 = (1.10 \pm 0.25) \times 10^{-6}\, r \cos \psi,
\eeq
  where $\psi$ is the angle between the direction of observation and the
  dipole axis, while the distance $r$ is measured in billions of light
  years. The confidence level of this result (as compared with a
  ``monopole'' model where values of $\alpha$ are the same in all
  directions) has been estimated as $4.1 \sigma$. A more detailed
  discussion of the observational data can be found, e.g., in \cite{webb3}.

  Let us also mention the laboratory experimental data on possible FPC
  variations in the modern epoch. The tightest constraints on $\alpha$
  variations have been obtained by comparison of readings of atomic clocks
  using optical transitions in Al and Hg ions (without using cesium clocks
  that have become classic) \cite{rosbnd1}:
  $(d\alpha/dt)/\alpha = (-1.6 \pm 2.3) \times 10^{-17}$ per year. This
  result is of the same order of magnitude as the tightest constraints
  obtained previously from an isotopic composition analysis of the decay
  products in the natural nuclear reactor that operated in the Oklo region
  (Gabon) about 2 billion years ago. Unlike the laboratory data, the Oklo
  results \cite{shlya} and, in particular, the tightest constraint
  \cite{Petrov,Gould}
\beq  		\label{a-Oklo}
        -3.7 \times 10^{-17}/{\rm yr} < d (\ln \alpha)/dt
                    < 3.1 \times 10^{-17}/{\rm yr}
\eeq
  rely on the assumption that during these 2 billion years the value of
  $\alpha$ changed uniformly, if changed at all. This assumption looks
  rather natural but actually follows from nowhere.

  Thus in the modern epoch, at least on Earth since the Oklo times, the
  parameter $\alpha$ did not change more rapidly than by approximately
  $10^{-17}$ per year. If, on the other hand, we use the distant quasar data
  and take a mean value of $d(\ln \alpha)/dt$ for about 10 billion years,
  we shall obtain a variation rate of about $10^{-15}$ per year. Therefore
  one can conclude that at times earlier (maybe much earlier) than 2 billion
  years ago the value of $\alpha$ changed relatively rapidly but afterwards
  stopped or almost stopped to change. The task of theory was to explain
  such a behavior; however, if one takes into account the most recent
  observations \cite{webb2, chiba}, one should add the necessity of
  exlaining the spatial variations of $\alpha$. Though, one cannot exclude
  the opportunity that the variations of $\alpha$ are purely spatial in
  nature whereas the time dependence is related to the finiteness of the
  velocity of light: being located at a fixed point and at fixed time,
  we receive signals from distant regions of the Universe emitted at earlier
  cosmological epochs, and it is therefore impossible to separate spatial
  and temporal dependences of the parameters.

  Let us briefly discuss the theoretical models describing variations of
  $\alpha$. Thus, following the pioneering ideas of Dirac and Eddington,
  Dicke and Peebles \cite{Dicke} in 1962 considered variations of $\alpha$
  in cosmological models admitting a variable gravitational interaction
  intensity. Staniukovich \cite{stan} in 1965 discussed different variants
  of combined FPC variations in connection with Dirac's Large Number
  Hypothesis. Bekenstein \cite{Bekenstein} in 1982 described a model of
  $\alpha$ variations on the basis of the most general assumptions on the
  electromagnetic interaction: covariance, gauge invariance, causality and
  invariance with respect to time reversion. This led to a modified Maxwell
  electrodynamics and provided a certain dynamics of $\alpha$.

  Since the advent of astronomical evidence on possible time variations of
  $\alpha$, there emerged a whole class of new models describing such
  variations by introducing certain scalar fields. Thus, Sandvik et al.
  \cite{Sandvik} proposed a cosmological extension of Bekenstein's theory
  \cite{Bekenstein} with a term of the form  $-\frac{1}{4} F\mn F\MN
  e^{-2\psi}$ in the initial Lagrangian, where the scalar field $\psi$
  interacts only with the electromagnetic field $F\MN$. The effect of the
  field $\psi$ in the dynamics of the expanding Universe was also considered.
  It was shown that in this model $\alpha$ remained almost constant in the
  radiation-dominated epoch, slightly increased in the matter-dominated
  epoch and approaches a constant value at times when the Universe expansion
  accelerates due to the presence of a positive cosmological constant.

  Spatial variations of $\alpha$ were also discussed much before they 
  were claimed to be really discovered, in attempts to explain the 
  discrepancy between cosmological and terrestrial data on $\alpha(t)$ 
  \cite{mota1, mota2}.

  In the recent attempts to explain both temporal and spatial variations of 
  the fine structure constant, quite popular are models assuming the 
  existence of domain walls connected with scalar field dynamics (see, e.g.,
  \cite{Chiba-1, Olive-1}). Thus, in \cite{Olive-1} the initial action
  contains a dilaton-like scalar field $\phi$ interacting with the
  electromagnetic field and having a potential of the form
  $V(\phi)=\frac{1}{4}\lambda(\phi^{2} - \eta^{2})^{2}$. A domain wall is
  formed due to spontaneous symmetry breakdown. At points separated by the
  domain wall the values of $\alpha$ are different, which can explaing the
  observable variations if the wall intersects our Hubble volume.

  In \cite{Odintsov} it has been shown that in $F(R)$ gravity it is
  possible to obtain a static solution in the form of an effective
  (gravitational) domain wall, and that the choice of a logarithmic
  nonminimal interaction of the electromagnetic field with gravity in the
  form
\[
     -\frac{1}{4}\left[1 + \ln\left(\frac{R}{R_0}\right)\right] F\mn F\MN
\]
  (where $R_0$ is the modern value of the scalar curvature) makes it
  possible to describe variations of $\alpha$, whose value grows as the
  curvature $R$ decreases.

  Olive et al. \cite{Olive-2} discuss a model with two domain walls,
  where the scalar field potential has three minima:
\[
    V(\phi) = \lambda \left(|\Phi|^2 - \frac{\eta^2}{2} \right)^2
    - \sqrt{2}i\epsilon\left(\Phi^3 - (\Phi^{*})^3 \right) + V_0
\]
  It turns out that such a model much better describes the observational
  data than a similar one \cite{Olive-1} with a single domain wall.

  The paper \cite{Barrow} suggests an extension of the previous
  BSBM (Bekenstein-Sandvik-Barrow-Magueijo) theory (\cite{Bekenstein,
  Sandvik}) by introducing a dependence of the coupling constant $\omega$
  of the scalar field $\psi$ on the field itself, so that the Lagrangian
  contains the terms
  $L_{\psi} = -\half \omega(\psi) \d_\mu\psi \d^\mu\psi$ and
  $L_{\rm em} = -\frac{1}{4} F\mn F\MN e^{-2\psi}$. The choice of
  $\omega(\psi)$ allows for obtaining both growing and falling time
  dependences of $\alpha$. This model differs from those with domain walls
  in that the variations of $\alpha$ are smooth and continuous, and a choice
  between these models must be easy with future more precise and reliable
  observational data.

  Mariano and Perivolaropoulos \cite{Mariano} have reported on a correlation
  between the spatial distribution of $\alpha$ values and the dipole
  anisotropy of the dark energy distribution. In the same paper they have
  suggested a theoretical model explaining this correlation (named
  ``extended topological quintessence'') which naturally predicts
  inhomogeneous spherical distributions of both the dark energy and the
  values of $\alpha$. The model assumes the existence of a huge global
  monopole with a size of Hubble order, which nonminimally interacts with
  the electromagnetic field. There emerge mutually related distributions of
  different parameters with a dipole anisotropy from the viewpoint of any
  observer located outside the monopole center. The monopole is formed after
  a phase transition in a set of three scalar fields with an O(3) symmetric
  Lagrangian.

  In a later paper \cite{Mari-2} the same authors support their inferences
  by the data on one more anisotropy also seeming to exist and to be aligned
  with other ``dipoles'', the so-called Large-Scale Velocity Flows (Dark
  Flow), i.e., recent indications that there is a large-scale peculiar
  velocity flow with an amplitude larger than 400 km/s on scales up to $100
  h^{-1}$ Mpc $(z \leq 0.03)$.

  It is also important to mention the theoretical models considering FPC
  variations in the framework of unification scenarios. In particular, P.
  Langacker et al. \cite{Langacker} consider possibile variations of
  coupling constants due to physics at very high energies, where the gauge
  couplings may be unified. It means that one should treat a joint variation
  of the fine structure and strong coupling constants.

  Similarly to \cite{Langacker}, X. Calmet and H. Fritzsch \cite{Calmet}
  discuss FPC variations in the context of Grand unification. They show that
  such a consideration leads to small time shifts of the nucleon mass, the
  magnetic moment of the nucleon and the weak coupling constant, and it is
  expected to have a relative change of the nucleon mass larger than that
  of $\alpha$ by a factor of $\sim 40$.

  A more detailed discussion of theoretical models involving unification
  scenarios to explain the FPC variations can be found in \cite{Uzan}.

  It should be noted that all the above approaches, to explain variations of
  $\alpha$, introduce scalar fields whose existence and manner of interaction
  with the electromagnetic field are postulated from the outset and are not
  explained in any way. In what follows, it will be shown how the scalar
  fields and their interaction law with electromagnetism naturally follow
  from nonlinear multidimensional gravity. Spatial variations of $\alpha$
  are explained by a large-scale inhomogeneity of this scalar field. The
  magnitude of this inhomogeneity is constrained by CMB observations
  \cite{WMAP}.                             ``

  The approach we are using has been formulated in \cite{VfromD}, where
  a methodology was suggested allowing for a transition from multidimensional
  gravity with higher derivatives to Einstein-Hilbert gravity with
  effective scalar fields. Later on this approach was successfully applied
  for a unified description of the inflationary stage of the Universe and the
  modern secondary inflation \cite{infl} and an explanation of the origin
  of the Higgs field \cite{Bolokhov:2010mz}; a mechanism of cascade reduction
  of multidimensional space to the observable one was suggested
  \cite{Rubin:2011jm, BRook}. It has been shown under which conditions the
  compact extra dimensions become stationary (i.e., have a constant volume),
  and the cause of their maximum symmetry was found \cite{KAAR}.

  The present study has been performed in the framework of this approach
  and is an example of its employment. The paper is organized as follows.
  Sec.\,2 briefly describes the general formalism used. In this framework,
  in Sec.\,3 we build a homogeneous and isotropic cosmological model able to
  describe the present accelerated Universe along with a time dependence
  of the fine structure constant $\alpha$. In Sec.\,4, this cosmological
  model is slightly perturbed on large scale, which enables us to explain
  spatial variations of $\alpha$.
  Sec.\,5 is a brief conclusion.

\section {Multidimensional gravity and its reduction}

  Consider a $(D = 4 + d_1)$-dimensional manifold with the metric
\beq                                                           \label{ds}
        ds^2 = g\mn dx^\mu dx^\nu + \e^{2\beta(x)} b_{ab} dx^a dx^b
\eeq
  where the extra-dimensional metric components $b_{ab}$ are independent of
  $x^{\mu}$, the observable four space-time coordinates.

  The $D$-dimensional Riemann tensor has the nonzero components
\bear
    R\MN{}_{\rho\sigma} \eql \oR\MN{}_{\rho\sigma},       \label{Riem}
\nn
    R^{\mu a}{}_{\nu a} \eql \delta^a_b\, B\nM, \cm
    B\nM := \e^{-\beta} \nabla_\nu (\e^\beta \beta^{\mu}),        
\nn
    R^{ab}{}_{cd} \eql
      \e^{-2\beta} \oR^{ab}{}_{cd} + \delta^{ab}{}_{cd} \beta_\mu\beta^\mu,
\ear
  where capital Latin indices cover all $D$ coordinates, the bar marks
  quantities obtained from $g\mn$ and $b_{ab}$ taken separately,
  $\beta_{\mu} \equiv \d_{\mu}\beta$ and $\delta^{ab}{}_{cd}\equiv
  \delta_{c}^{a}\delta_{d}^{b}-\delta_{d}^{a}\delta_{c}^{b}$. The
  nonzero components of the Ricci tensor and the scalar curvature are
\bear
    R\mN \eql \oR\mN + d_1\, B\mN,          \label{Ric}           
\nnv
    R_a^b \eql \e^{-2\beta} \oR_a^b
                    + \delta_a^b [ \Box \beta + d_1 (\d{\beta})^2 ],
\nnv
    R \eql \oR [g] + \e^{-2\beta }\oR [b] + 2d_1 \Box \beta
           + d_1(d_1+1) (\d{\beta})^2,
\ear
  where $(\d{\beta})^2 \equiv \beta_{\mu}\beta^{\mu}$,
  $\Box = \nabla^\mu \nabla_\mu$ is the d'Alembert operator while $\oR[g]$
  and $\oR [b]$ are the Ricci scalars corresponding to $g\mn$ and $b_{ab}$,
  respectively. Let us also present, using similar notations, the
  expressions for two more curvature invariants, the Ricci tensor squared
  and the Kretschmann scalar $\cK = R^{ABCD}R_{ABCD}$:
\bearr
    R_{AB}R^{AB} = \oR\mn\oR\MN + 2d_1 \oR \mn B\MN + d_1^2 B\mn B\MN
        + \e^{-4\beta}\oR_{ab}\oR^{ab}
\nnn\inch
    + 2\e^{-2\beta} \oR[b] [\Box\beta + d_1 (\d{\beta})^2 ]
      + d_1 [\Box\beta + d_1 (\d{\beta})^2 ]^2,             \label{Ric2}
\yyy
    \cK = \overline{\cK}[g] + 4 d_1 B\mn B\MN + \e^{-4\beta}\overline{\cK}[b]
      + 4 \e^{-2\beta} \oR [b] (\d{\beta})^2
         +2 d_1 (d_1-1) [(\d\beta)^2 ]^2 .                     \label{Kre}
\ear

  Suppose now that $b_{ab}$ describes a compact $d_1$-dimensional space of
  nonzero constant curvature, i.e., a sphere ($K=1$) or a compact
  $d_1$-dimensional hyperbolic space ($K = -1$) with a
  fixed curvature radius $r_0$ normalized to the $D$-dimensional
  analogue $\mD$ of the Planck mass, i.e., $r_0 = 1/\mD$ (we use the
  natural units, with the speed of light $c$ and Planck's constant $\hbar$
  equal to unity). We have
\bear
    \oR^{ab}{}_{cd} \eql K\,\mD^2\,\delta^{ab}{}_{cd},         \label{r0}
\nn
    \oR_a^b \eql K\,\mD^2\, (d_1-1) \delta_a^b,
\nn
    \oR [b] \eql K\,\mD^2\, d_1 (d_1-1) = R_b.
\ear
  The scale factor $b(x) \equiv \e^{\beta}$ in (\ref{ds}) is thus kept
  dimensionless; $R_b$ has the meaning of a characteristic curvature scale
  of the extra dimensions.

  Consider, in the above geometry, a sufficiently general
  curvature-nonlinear theory of gravity with the action
\bear                                                         \label{act1}
     S \eql \Half \mD^{D-2} \int\sqrt{^{D}g}\,d^{D}x\,(L_g + L_m),
\nn
     L_g \eql F(R) + c_1 R^{AB}R_{AB} + c_2 \cK,
\ear
  where $F(R)$ is an arbitrary smooth function, $c_1$ and $c_2$ are
  constants, $L_m$ is a matter Lagrangian and ${^D}g = |\det(g_{MN})|$.

  The extra coordinates are easily integrated out, reducing the action
  to four dimensions:
\beq
     S = \Half \cV [d_1]\,\mD^2 \int\sqrt{^4g}\,d^{4}x\,
            \e^{d_1\beta}\,[L_g + L_m],                      \label{act2}
\eeq
  where $^{4}g = |\det(g\mn)|$ and $\cV[d_1 ]$ is the volume of a
  compact $d_1$-dimensional space of unit curvature.

  \eq (\ref{act2}) describes a curvature-nonlinear theory with non-minimal
  coupling between the effective scalar field $\beta$ and the curvature.
  Let us simplify it in the following way (putting, for convenience,
  $\mD=1$, so that all quantities are now expressed in ($D$-dimensional)
  Planck units:

\medskip\noi
{\bf (a)} Express everything in terms of 4D
    variables and $\beta(x)$; we have, in particular,
\bearr                                                      \label{R4}
        R = R_4 + \phi + f_1, \cm
        R_4 = \oR [g],      \cm
            f_1 = 2d_1 \Box \beta + d_1(d_1+1)(\d{\beta})^2,
\ear
    where we have introduced the effective scalar field
\beq                                                         \label{phi}
        \phi (x) = R_b \e^{-2\beta (x)}
              = K d_1(d_1-1)\, \e^{-2\beta (x)}   
\eeq
    The sign of $\phi$ coincides with $k = \pm 1$, the sign of curvature in
    the $d_1$ extra dimensions.

\medskip\noi
{\bf (b)} Suppose that all quantities are slowly varying, i.e., consider
    each derivative $\d_{\mu}$ (including those in the definition of $\oR$)
    as an expression containing a small parameter $\eps$; neglect all
    quantities of orders higher than $O(\eps^2)$ (see \cite{VfromD, BRook}).

\medskip\noi
{\bf (c)} Perform a conformal mapping leading to the Einstein conformal
    frame, where the 4-curvature appears to be minimally coupled to the
    scalar $\phi$.

\medskip
  In the decomposition (\ref{R4}), both terms $f_1$ and $R_4$ are regarded
  small in our approach, which actually means that all quantities,
  including the 4D curvature, are small as compared with the
  $D$-dimensional Planck scale. The only term which is not small is
  $\phi$, and we can use a Taylor decomposition of the function $F(R) =
  F(\phi + R_4 + f_1)$:
\bearr                                                       \label{Fapprox}
    F(R) = F(\phi + R_4 + f_1 )
    \simeq F(\phi) + F'(\phi)\cdot(R_4 +f_1 )+...,
\ear
  with $F'(\phi)\equiv dF/d\phi$. Substituting this, and the corresponding
  decompositions of the expressions (\ref{Ric2}) and (\ref{Kre}), into \eq
  (\ref{act2}), we obtain, up to $O(\eps^2)$, the following effective
  gravitational Lagrangian $L_g$ in \eq (\ref{act2}):
\bearr
      L_g = F'(\phi) R_4  + F(\phi) + F'(\phi) f_1 + c_*\phi^2
      + 2 c_1\phi \Box\beta
            + 2(c_1 d_1 + 2c_2) (\d\beta)^2                \label{Lg_4}
\ear
  with  $c_* = c_1/d_1 + 2c_2/[d_1(d_1-1)]$.

  The action (\ref{act2}) with (\ref{Lg_4}) is typical of a scalar-tensor
  theory (STT) of gravity in a Jordan frame. To study the dynamics of the
  system, it is helpful to pass on to the Einstein frame. Applying the
  conformal mapping
\bearr  \nhq                                                \label{trans-g}
    g\mn \ \mapsto \tg\mn = |f(\phi)| g\mn,
\qquad
            f(\phi) =  \e^{d_1\beta}F'(\phi),
\ear
  after a lengthy calculation, we obtain the action in the Einstein frame as
\bear
     S \eql \Half \cV[d_1] \int \sqrt{\tg}\, (\sign F') L,
\nn
     L \eql \tR_4 + K_{\rm E}(\phi) (\d\phi)^2
                        - 2V_{\rm E}(\phi) + {\wt L}_m,      \label{Lgen}
\\
     {\wt L}_m \eql (\sign F')\frac{\e^{-d_1\beta}}{F'(\phi)^2} L_m;
                                 \label{Lm}
\\ \nq
      K_{\rm E}(\phi) \eql                                   \label{KE}
        \frac{1}{4\phi^2} \biggl[
            6\phi^2 \biggl(\frac{F''}{F'}\biggr)^2\!
            -2 d_1 \phi \frac{F''}{F'}
        + \Half d_1 (d_1{+}2) + \frac{4(c_1 + c_2)\phi}{F'}\biggr],
\\ \nq
       -2V_{\rm E}(\phi) \eql (\sign F') \frac{\e^{-d_1\beta}}{F'(\phi)^2}
                [F(\phi) + c_* \phi^2],
                                  \label{VE}
\ear
  where the tilde marks quantities obtained from or with $\tg\mn$; the
  indices are raised and lowered with $\tg\mn$; everywhere $F = F(\phi)$ and
  $F' = dF/d\phi$; $\e^{\beta}$ is expressed in terms of $\phi$ using
  (\ref{phi}).

  Let us consider the electromagnetic field $F\mn$ as matter in the initial
  Lagrangian, putting
\beq                                                        \label{Lem-1}
               L_m = \alpha_1^{-1} F\mn F\MN,
\eeq
  where $\alpha_1$ is a constant. After reduction to four dimensions this
  expression acquires the factor $\e^{d_1\beta}$ arising from the metric
  determinant: $\sqrt{^D g} = \sqrt{^4 g} \e^{d_1\beta}$. In the subsequent
  transition to the Einstein picture the expression $\sqrt{^4 g}F\mn F\MN $
  remains the same (it is the well-known conformal invariance of the
  electromagnetic field), hence the Lagrangian (\ref{Lm}) takes the form
\beq
             {\wt L}_m = \alpha_1^{-1} \e^{d_1\beta} F\mn F\MN,
\eeq
  and for the effective fine structure constant $\alpha$ we obtain
\beq                                                         \label{var-A}
             \frac{\alpha}{\alpha_0} = \e^{d_1(\beta_0 - \beta)},
\eeq
  where $\alpha_0$ and $\beta_0$ are values of the respective quantities
  at a fixed space-time point, for instance, where and when the observation
  is taking place.

\section {The cosmological model}
\def\od{{\overline d}}

  Depending on the choice of $F(R)$, the parameter $c_1$ and $c_2$ and the
  matter Lagrangian in the action (\ref{act1}), the theory under consideration
  can lead to a great variety of cosmological models. Some of them were
  discussed in \cite{VfromD}, mostly those related to minima of the
  effective potential (\ref{VE}) at nonzero values of $\phi$. Such minima
  correspond to stationary states of the scalar $\phi$ and consequently of
  the scale factor $b = \e^{\beta}$ of the extra dimensions. If the minimum
  value of the potential is positive, it can play the role of a cosmological
  constant that launches an accelerated expansion of the Universe.

  Here, we would like to focus on another minimum of the potential
  $V_{\rm Ein}$, existing for generic choices of the function $F(R)$ with
  $F' >0$ and located at the point $\phi =0$. The asymptotic $\phi \to 0$
  corresponds to growing rather than stabilized extra dimensions: $b =
  \e^{\beta}\sim 1/\sqrt{|\phi|} \to \infty$. A model with such an
  asymptotic growth at late times may still be of interest if the growth is
  sufficiently slow and the size $b$ does not reach detectable values by
  now. Let us recall that the admissible range of such growth comprises as
  many as 16 orders of magnitudes if the $D$-dimensional Planck length
  $1/\mD$ coincides with the 4D one, i.e., about $10^{-33}$ cm: the upper
  bound corresponds to lengths about $10^{-17}$ cm or energies of the order
  of a few TeV. This estimate certainly changes if there is no such
  coincidence.

  One should note that small values of $\phi$ to be considered here are
  still very large as compared to 4D quantities, and so our general
  assumptions are well justified. Indeed, according to (\ref{phi}),
\[
       |\phi| = \frac{d_1(d_1 - 1)}{b^2},
\]
  where $b \lesssim 10^{16}$, hence $|\phi| \gtrsim d_1^2\cdot 10^{-32}$,
  while the quantity $\tR_4$, if identified with the curvature of the
  modern Universe, is of the order $10^{-122}$ in Planck units (that is,
  close to the Hubble parameter squared, or (the Hubble time)${}^{-2}$, see
  also \eq (\ref{Hubble}) below).

  Let us check whether it is possible to describe the modern state of
  the Universe by an asymptotic form of the solution for $\phi\to 0$
  as a spatially flat cosmology with the 4D Einstein-frame metric
\beq
        d{\wt s}{}^2_4 = dt^2 - a^2 (t) d\vec x{}^2,         \label{dsE}
\eeq
  where $a(t)$ is the Einstein-frame scale factor. We shall be are working
  in the framework of quadratic gravity with a cosmological constant,
  i.e.,\footnote
      {We assume for certainty $\phi > 0$, or, which is the same according
       to (\ref{phi}), $K = +1$, but everything can be easily reformulated
       for $\phi < 0$.}
\beq                                                         \label{F_0}
      F(\phi) = -2\Lambda_D + F_2\phi^2 ,
\eeq
  where $\Lambda_D$ is the initial cosmological constant. Then, substituting
  $F' =2\phi$ and $F'' =2$, we obtain for the kinetic and potential terms in
  the Lagrangian (\ref{Lgen}) in the first approximation in $\phi$:
\bearr
        K_{\rm E} \approx K_0/(2\phi^2),
\cm                                                        \label{KV_0}
        K_0 = \Half \biggl[\Half d_1^2 - d_1 + 6 + 2(c_1+c_2)\biggr];
\nnn
        V_{\rm E} \approx V_0 \e^{-2\od \beta},
\cm
        V_0 = \frac{\Lambda_D}{4 d_1^2(d_1-1)^2}, \cm 2\od = d_1-4.
\ear
  It is clear that this model can work only if $d_1 > 4$.
  In terms of $\beta$ instead of $\phi$, the Lagrangian takes the form
\beq
    L = \tR_4 + 2K_0 (\d\beta)^2
                       - 2V_0 \e^{-2\od \beta}+ {\wt L}_m,    \label{L-bet}
\eeq
  Neglecting the gravitational influence of the electromagnetic field
  (that is, considering only vacuum models), one can write down the
  independent components of the Einstein and scalar field equations
  with the unknowns $\beta(t)$ and $a(t)$ in the form
\bear
       3\frac{\dot a{}^2}{a^2}                                \label{eq-a4}
            \eql K_0 \dot\beta{}^2 + V_0 \e^{-2\od\beta},
\\                                                            \label{eq-b4}
      \ddot \beta + 3\frac{\dot a}{a} \dot{\beta}
                \eql \frac{V_0 \od} {K_0} \e^{-2\od\beta}.
\ear

  These equations, corresponding to a scalar field with an exponential
  potential, can be solved exactly but the solution looks rather involved,
  and for our purpose more preferable is the comparatively simple approximate
  solution that can be obtained in the slow-rolling approximation; the
  latter should be acceptable at late times. Let us suppose that
\beq                                                       \label{rolling}
       |\ddot \beta| \ll 3\frac{\dot a}{a} \dot {\beta},
   \qquad
        K_0 \dot\beta{}^2 \ll V_0 \e^{-2\od\beta},
\eeq
  and neglect the corresponding terms in \eqs (\ref{eq-a4}) and
  (\ref{eq-b4}). Then, expressing the quantity $\dot a/a$ from (\ref{eq-a4})
  and substituting it into (\ref{eq-b4}), we obtain
\beq                                                           \label{eq-b}
    \dot\beta = \frac{\od\sqrt{V_0}}{K_0 \sqrt{3}}\e^{-\od\beta},
\eeq
  whence
\beq                                                           \label{sol-b}
    \e^{\od\beta} = \frac{\od^2}{K_0}\sqrt{\frac{V_0}3}(t+t_1),
\eeq
  where $t_1$ is an integration constant. For the scale factor $a(t)$ we have
\beq                                                        \label{sol-a}
     \frac {\dot a} a = \frac p {t+t_1} \quad\then\quad
        a =  a_1 (t+t_1)^p,  \qquad
        a_1 = \const,   \qquad  p = \frac {K_0}{\od^2}.
\eeq

  Substituting the solution to the slow-rolling conditions (\ref{rolling}),
  we make sure that they hold as long as $p \gg 1$, or in terms of
  the input parameters of the theory,
\beq                                                         \label{roll}
       p = \frac{d_1^2 -2 d_1 +12 + 4(c_1+c_2)}{(d_1-4)^2} \gg 1.
\eeq
  We will assume that this condition holds.

  A further interpretation of the results depends on which conformal
  frame is regarded physical (observational) \cite{bm-predict,erice},
  and this in turn depends on the manner in which fermions appear in the
  (so far unknown) underlying unification theory involving all interactions.

  Let us adopt the simplest hypothesis that the observational picture
  coincides with the Einstein picture and make some estimates. Thus, the
  inverse of the modern value of the Hubble parameter (the Hubble time)
  is estimated as
\beq                                                        \label{Hubble}
    t_H = 1/H_0 = a_0/{\dot a}_0 \approx 4,4 \times 10^{17} {\rm c}
                \approx 8 \times 10^{60}\, t_{\rm pl},
\eeq
  where $t_{\rm pl}$ is the Planck time and the index ``0'' marks quantities
  belonging to the present time, which is a usual notation in cosmology.
  From (\ref{sol-a}) it follows that $H_0 = p/(t_0 + t_1)$, whence
\beq
       t_* := t_0 + t_1 = p t_H \gg t_H.
\eeq
  With $p \gg 1$, the model satisfies the observational constraints on the
  factor $w$ in the effective equation of state $p = w\rho$ of dark energy
  that causes the accelerated expansion of the Universe: at $w = \const$ we
  have $a \sim t^{2/(3+3w)}$, consequently, $w = -1 + 2/(3p)$ is a number
  close to $-1$: for example, to have $w \approx -0.99$, one should put only
  $p = 66$. Meanwhile, the recent observational data allow for a
  comparatively large range of $w$ \cite{w1,w2,w3,w4} but anyway admitting
  $w=-1$ corresponding to a cosmological constant. This follows from
  combining the recent measurements of cosmic microwave background
  anisotropies, Supernovae luminosity distances, baryonic acoustic
  oscillations, and $H(z)$ measurements, though different tests lead to
  different confidence intervals.

  Furthermore, the ``internal'' scale factor $b(t)= \e^\beta$ grows much
  slower than $a(t)$:
\beq                                                          \label{b_t}
       b(t) = b_0 \biggl(\frac{t+t_1}{t_*}\biggr)^{1/\od},
    \cm
          b_0 = \biggl(\frac 1{H_0} \sqrt{\frac {V_0}{3}}\biggr)^{1/\od}.
\eeq
  Using the expression for $V_0$ from (\ref{KV_0}), one can estimate the
  initial parameter $\Lambda_D$, connecting it with the size of the extra
  factor space $b_0$: in Planck units,
\beq                                                  \label{Lam-est}
       \Lambda_D = 12 H_0^2 d_1^2 (d_1-1)^2 b_0^{d_1-4}
        \approx \frac {3}{16} d_1^2 (d_1-1)^2 b_0^{d_1-4} \times 10^{-120}.
\eeq
  As already mentioned, the ``internal'' scale factor $b= \e^\beta$
  should be in the range $1 \ll b_0 \lesssim 10^{16}$ in Planck units.
  The estimate (\ref{Lam-est}) shows that the present model makes much
  easier the well-known ``cosmological constant problem'' (the difficulty of
  explaining why in standard cosmology $\Lambda_{\rm standard} \sim
  10^{-122}$ in Planck units). For instance, if (in the admissible range)
  $b_0 = 10^{15}$ and $d_1 = 12$, it follows $\Lambda_D = 3267$ without
  any indication of fine tuning.

  Let us estimate the possible range of the parameters $c_1$ and $c_2$ in
  the action (\ref{act1}). The present model describes only the modern stage
  of the Universe evolution, but it should admit an improvement after which
  it will account for other stages, including the early inflation. Then one
  should require that the curvature-nonlinear terms in the initial
  Lagrangian should not violate our slow-change approximation, see Sec.\,2
  This leads to the condition $c_{1,2}\ll 10^{11}$. Indeed, during
  inflation, the Hubble parameter is $H\sim 10^{-6}$ in Planck units, while
  the scalar curvature at inflation, when the 4D geometry is approximately
  de Sitter, is estimated as $R \simeq 12H^2 \sim 10^{-11}$. Assuming that
  the Ricci and Riemann tensor components have the same order of magnitude,
  we find that the condition $R \gg c_1 R^{AB}R_{AB}$, used above in the
  framework of the slow-change approximation, will be violated if $c_1$ is
  too large. The upper bound of the parameter $c_2$ is obtained in a similar
  way.

  The smallness of the observed variations of $\alpha$ leads to
  another constraint on $c_1$ and $c_2$: according to (\ref{var-A}),
\beq                                                        \label{var-A}
    \alpha/\alpha_0 = (b/b_0)^{-d_1}
        = \biggl(\frac{t+t_1}{t_0+t_1}\biggr)^{-2d_1/(d_1-4)}
                \approx 1 - \frac{2d_1}{d_1-4}\frac {t-t_0}{t_*},
\eeq
  so that $\dot \alpha/\alpha \sim 10^{-10}/p$ per year. By the empirical
  data, this quantity cannot be larger than about $10^{-17}$ per year.
  A comparison leads to the constraint $p \gtrsim 10^7$ and hence the
  effective equation-of-state parameter $w$ is equal to $-1$ up to seven
  meaningful digits. Taking into account the relation (\ref{roll}) between
  $p$ and the input parameters $c_1$ and $c_2$, we obtain similar bounds on
  these parameters if the number of extra dimensions $d_1$ is not too large.

  Thus the allowed range of $c_1$ and $c_2$ (assuming that they are of the
  same order of magnitude),
\beq                                \label{ineq1}
        10^7 \lesssim  c_{1,2} \ll 10^{11}
\eeq
  is wide enough, which means that any fine tuning is absent.

  One of the well-known constraints on Kaluza-Klein-like cosmologies is the
  requirement of a sufficiently slow evolution of the internal scale factor,
  e.g., according to \cite{arkani98}, the extra-dimensional volume should
  not have changed by more than 10\,\% since the times of primordial
  nucleosynthesis. This requirement rests on the relation between the
  effective Newtonian gravitational constant $G_{\rm N}$ and the volume of
  extra dimensions. However, in the Einstein conformal frame used here,
  $G_{\rm N} = \const$ by definition, therefore the above constraint does
  not apply here. Still, even if we passed on to the Jordan frame, the
  relative variations of $G_{\rm N}$ would be the same as those of $\alpha$
  and obey the law (\ref{var-A}), i.e., within about $10^{-5}$ for the Hubble
  time.

  In the next section we shall see that the inequality $p \gtrsim 10^7$
  and consequently $c_{1,2} \gtrsim 10^7$ are substantially relaxed in the
  perturbed model.

\section {Spatial variations of $\alpha$}

  In the previous section we discussed the properties of a homogeneous model
  which does not contain any spatial variation of $\alpha$ (and any other
  physical quantity). Let us try to describe variations of $\alpha$ by taking
  into account spatial perturbations of the scalar field and the metric.
  Only long-wave perturbations will be of interest for us, with
  characteristic lengths of the order of the horizon size.

  An observed statistically isotropic sky means that there is no preferred
  axis. Nevertheless, super-horizon components were produced by quantum
  fluctuations at the beginning of inflation in the same way as fluctuations
  of smaller scale. It means that the dipole component must exist though
  hardly observed due to its contamination by the Doppler effect caused by
  the motion of our Local Group with respect to the CMB. Hence there must
  exist a weakly expressed distinguished direction along which the metric
  and scalar field inhomogeneity is most clearly pronounced.

  Accordingly, we now choose a metric more general than (\ref{dsE}),
\beq                                                         \label{ds1}
     ds_{\rm E}^2 =
    \e^{2\delta\gamma}dt^2 - a(t)^2\e^{2\delta\lambda}dx^2
    				-a(t)^2 \e^{2\delta\eta}(dy^2 + dz^2),
\eeq
  where $x$ is the distinguished direction and $\delta\gamma, \delta\lambda,
  \delta\eta \ll 1$ are functions of $x$ and $t$. In addition, we replace
  the effective scalar field $\beta (t)$ with $\beta(t) + \delta\beta(x,t)$.

  Then the relevant Einstein-scalar equations corresponding to the Lagrangian
  (\ref{L-bet})
  can be written as follows (preserving only terms linear in the ``deltas''):
\bearr
    \delta\ddot{\beta} + \frac{3\dot a}{a} \delta\dot{\beta}
    + \dot\beta (\delta\dot\lambda - \delta\dot \gamma)
    - \frac{1}{a^2} \delta \beta''                       \label{sc-d}
    + \frac{1}{2K_0} \delta(V_\beta \e^{2\gamma}) =0,
\\ \lal                                                         \label{22-d}
     \frac{\dot a}{a}(\delta\dot\lambda - \delta\dot \gamma)
            = \delta(V \e^{2\gamma}),
\\ \lal                                                         \label{01-d}
     \frac{\dot a}{a} \delta\gamma' = K_0 \dot\beta \, \delta\beta',
\ear
  where we have chosen the gauge (in other words, the reference frame in
  perturbed space-time) $\delta \eta \equiv 0$, the dot and the prime stand
  for $\d/\d t$ and $\d/\d x$, respectively. We have also denoted
  $V = V_{\rm E} = V_0 \e^{-2\od \beta}$ and $V_\beta = dV/ d\beta$.

  Integration of (\ref{01-d}), without loss of generality, leads to
\beq
        \delta\gamma = \frac{K_0}{H} \dot{\beta} \delta\beta,
\eeq
  where, as before, $H = \dot a/a$. This equation enables us to estimate
  the quantity $\delta\beta$. Indeed, according to the CMB data \cite{WMAP},
  we can take $\delta\gamma \sim 10^{-5}$, while the coefficient before
  $\delta\beta$ is of the order of unity at the present epoch
  (according to \eq (\ref{sol-a}) we have $(K_0/H_0)\dot{\beta}(t = t_0) =
  K_0/(\od \cdot p) = \od \sim 1$), we obtain $\delta\beta \sim 10^{-5}$.

  Substituting this $\delta\gamma$ to (\ref{sc-d}) and taking the difference
  $\delta\dot\lambda - \delta\dot \gamma$ from (\ref{22-d}), we finally
  arrive at the following single wave equation for $\delta\beta$:
\beq                                                         \label{eq-db}
    \delta \ddot \beta + \frac{3\dot a}{a} \delta\dot\beta
        -\frac{1}{a^2} \delta\beta''
     + \delta\beta \biggl[ \frac{2 \dot\beta{}^2}{H^2} VK_0
    + \frac{2 \dot\beta}{H} V_\beta + \frac 1{2K_0}V_{\beta\beta}\bigg]
        =0.
\eeq
  with an arbitrary constant $K_0$ and an arbitrary potential $V (\beta)$.
  In our case, with $V = V_0 \e^{-2\od \beta}$ and $K_0$ given in
  (\ref{KV_0}), we obtain
\beq                                                         \label{eq-db1}
    \delta \ddot \beta + \frac{3\dot a}{a} \delta\dot\beta
        -\frac{1}{a^2} \delta\beta''
         + \frac {2V_0 \e^{-2\od \beta}} {p}\delta\beta =0,
\eeq
  while the background quantities $a(t)$ and $\beta(t)$ are determined by
  the solution (\ref{sol-b}), (\ref{sol-a}). It remains to find a solution
  for $\delta\beta$ which, being added to the background $\beta(t)$, would
  be able to account for the observed picture of variations of $\alpha$.

  Since the background is $x$-independent, we can separate the
  variables and assume
\[
    \delta\beta = y(t) \sin k(x+x_0)
\]
  where $k$ has the meaning of a wave number, of order of the cosmological
  horizon scale, and $y(t)$ must be as small as $10^{-5}$. Then $y(t)$ obeys
  the equation
\beq                                                          \label{eq-y}
    \ddot y + \frac{3p}{t+t_1} \dot y
        + \biggl[ \frac{k^2}{a_1^2(t+t_1)^{2p}}
        + \frac {6p}{(t+t_1)^2}\biggr]y = 0.
\eeq
  Since the equation (\ref{eq-y}) has been derived in a certain approximation
  and describes only a restricted period of time close to the present epoch,
  it is reasonable to seek the solution in the form of a Taylor series:
\beq
    y(t) = y_0 + y_1 (t-t_0) + \Half y_2 (t-t_0)^2 + \ldots,
    \qquad y_i = \const.
\eeq
  Then $y_0$ and $y_1$ can be fixed at will as initial conditions, and \eq
  (\ref{eq-y}) leads to expressions of $y_2,\,y_3,\ldots$ in terms of
  $y_0$ and $y_1$. Even more than that, for a certain neighborhood of
  $t=t_0$ we can simply suppose $y = y_0 + y_1 (t-t_0)$. Actually, this
  approximation is good enough for $t-t_0 \ll t_* = t_0 -t_1$.

  In this approximation we obtain the following expression for
  variations of $\alpha$:
\beq                                                      \label{dA1}
      \frac{\alpha}{\alpha_0} \approx 1 - \frac{d_1}{\od}\,\frac{t-t_0}{t_*}
            - d_1 \sin[k(x+x_0)]\,[y_0 + y_1(t-t_0)] + O(\ep^2),
\eeq
  where $O(\ep^2)$ means $O((t-t_0)^2/t_*^2)$. Assuming that the observer is
  located at $x=0$ and requiring $\alpha/\alpha_0 = 1 + O(\ep^2)$ at $x=0$,
  we obtain the condition
\beq
     y_1 \sin (kx_0) = - 1/(\od \,t_*).                     \label{y_1}
\eeq
  This explains very small, if any, variations of $\alpha$ on Earth at
  present and since the Oklo times. Indeed, since $2\ten{9}\,{\rm yr}
  \approx \frac 17 t_H$ while $t_* = pt_H$, the addition $O(\ep^2)$ is of
  the order of $1/(50 p^2)$, where $p \gg 1$. If we take, for instance,
  $p=1000$, then at the Oklo time ($2\ten{9}$ years ago) we obtain
  a relative $\alpha$ variation of the order $0.5\ten{-8}$, which makes
  about $0.25 \ten{-17}$ per year.

  A substitution of (\ref{y_1}) and (\ref{y_1}) into (\ref{dA1}) at
  $t-t_0 = -x$ for $x > 0$ gives
\beq
       \alpha/\alpha_0 \approx 1 - d_1 y_0 \sin (kx_0)     \label{dA2}
                   + d_1 y_0 \,kx \cos (kx_0) + O(\ep^2)
\eeq
  at $x \ll t_*$. The same result is obtained if we substitute
  $t-t_0 = x$ for $x < 0$.

\begin{figure}[h]
\centering
\includegraphics[scale=0.6]{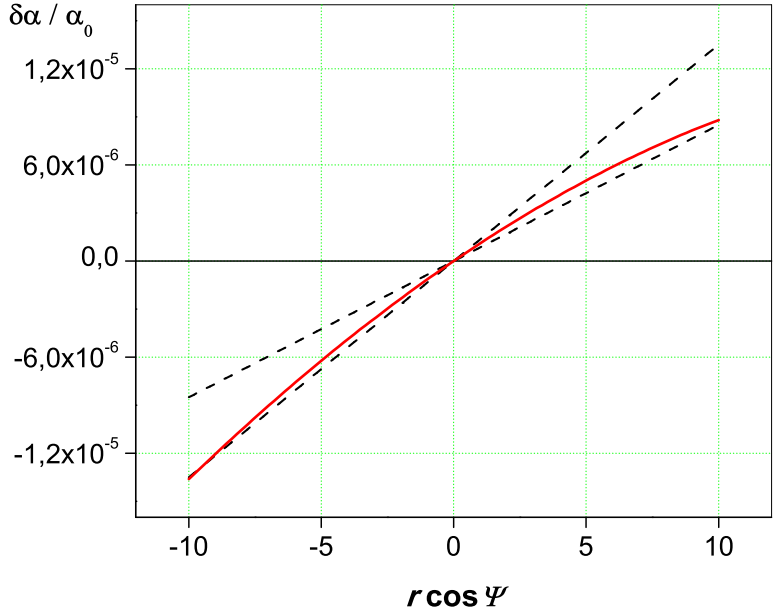}
\caption{The $r$ dependence of $\delta\alpha/\alpha_0$ (the distance
	$r$ is measured in billions of light years). The dashed lines
	correspond to \eq (\ref{dipole}), the solid red line to \eq
	(\ref{dA1}) at the parameter values $d_1 = 12$, $p = 10^{7}$, $y_0 =
	-4.7\ten{-6}$, $y_1 = -10^{-7}$ (bill. years)$^{-1}$, $k = 0.02$
	(bill. light years)$^{-1}$, $x_0 = 1$ billion of light years.}
    \label{picture}
\end{figure}

  Fig.\,\ref{picture} compares the observational data and the predictions
  of our model with the parameters indicated there. Recall that we are
  considering long-wave fluctuations, such that $k\leq 1/r_H \sim 0.1$
  (billion years)$^{-1}$ (where $r_H$ is the modern horizon size), with
  a small magnitude $y_0 \leq 10^{-5}$. The relations obtained are in
  good agreement with these estimates.
  We are using the conventional normalization $a_0 = 1$.

  Evidently, our model, in addition to the input theoretical parameters like
  $d_1,\, c_1,\ c_2$, contains the parameters $k$, $x_0$, $y_0$, $y_1$,
  depending on the initial form of the extra space metric.

  In the framework of chaotic inflation, these parameters vary in different
  regions of the visible part of the Universe. Their choice enables us to
  explain the spatial variations of $\alpha$ in agreement with the
  observations \cite{webb2}. Actually, there are only two conditions imposed
  on them: (\ref{y_1}) and the relationship identifying (\ref{dA2}) with
  the expression (\ref{dipole}) at $r=x$ and $\cos\psi =1$, i.e., on the
  dipole axis. We obtain (in Planck units)
\beq                                                       \label{x_0}
        d_1 y_0 k \cos (kx_0) \approx -2 \times 10^{-66}.
\eeq
  (This numerical value is used for obtaining the solid line in Fig.\,1.)
  The small constant shift of the $\alpha$ value at $x=0$ against the
  background does not change the interpretation of the results obtained.
  It should be stressed that (\ref{x_0}) is not a fine-tuning relation
  but simply fitting of the model parameters to the observational data. In
  fact, the very small number in the r.h.s. of (\ref{x_0}) results from the
  natural scale of $k \sim 1/r_H$, where $r_H \sim 10^{28}\ {\rm cm} \sim
  10^{61}\,\ell_{\rm pl}$ is the Hubble radius in terms of the Planck length
  $\ell_{\rm pl}$; five more orders of magnitude in (\ref{x_0}) are related
  to the smallness of $\alpha$ variations.

  The input parameters $c_1$ and $c_2$ are now not so strongly constrained
  by the condition of slow variations of $\alpha$ on Earth: this condition
  is already provided by the equality (\ref{y_1}) if we take $p \gtrsim
  1000$, hence $c_1 + c_2 \gtrsim 1000$. The approximation $p\gg 1$, in
  which our solution has been obtained, then also holds quite well. The
  inequality (\ref{ineq1}) is thus replaced by a much weaker one:
\beq                                              \label{ineq2}
        1000 \lesssim c_1 \sim c_2 \ll 10^{11}.
\eeq

\section{Conclusion}

  We have studied the possible effect of extra dimensions on large-scale
  variations of the fine structure constant $\alpha$ in space and time.
  In the multidimensional paradigm under consideration, the observable
  values of $\alpha$ and probably other physical quantities, including
  fundamental constants, depend on the size of the extra factor space.
  Variations of the dark energy density can be mentioned as an example.
  Indeed, the space-time variations of the energy density are dominated
  by those of the potential $V = V_{\rm E}$ given in (\ref{KV_0}).
  The relative variation $\delta V/V = -2\od\delta\beta$ is of the same
  order of magnitude as the space-time variations of $\alpha$ according
  to (\ref{var-A}). They are too small to be observed in the near future.

  We have discussed the dipole component only, but it seems evident that
  the same basis is applicable to higher multipoles in $\alpha$ variations.
  It means that the observational data, being quite uncertain, ``feel''
  these components, and further observations may detect them.

  We have focused on the behavior of $\alpha$ because it is the only
  fundamental constant for which there are more or less reliable data
  indicating its variations. We are also planning to analyze the behavior of
  other constants, above all, the gravitational constant and the particle
  masses.

  The model described here does not consistently include other kinds of
  matter than dark energy (represented by a scalar field of multidimensional
  origin). However, even such a simple model shows an agreement with the
  observational data (see Fig.\,1). The same numerical parameters also well
  agree with the CMB constraints which impose an upper bound on the
  fluctuation magnitude of the extra-dimensional metric.

  An advantage of the present model of $\alpha$ variation against many
  others (e.g., \cite{Chiba-1, Olive-1, Olive-2}) is that it assumes a
  common origin of dark energy and FPC variations. Our model also
  predicts the existence of higher multipoles in $\alpha$ variations.

\subsection*{Acknowledgments}

  The authors wish to thank A. Panov for his interest in our work. The work
  of S.R. and I.S. was supported in part by the Ministry of Education and
  Science of the Russian Federation, project 14.A18.21.0789.

\end{document}